# Automatic Detection of Microaneurysms in OCT Images Using Bag of Features

Elahe Sadat Kazemi Nasab[1,2], Ramin Almasi[3], Bijan Shoushtarian[1], Ehsan Golkar[2], Hossein Rabbani[2]*

1. Department of Artificial Intelligence, Faculty of Computer Engineering, University of Isfahan, Isfahan, Iran

2. Medical Image & Signal Processing Research Center, Isfahan University of Medical Sciences, Isfahan, Iran

3. Department of Computer Architecture, Faculty of Computer Engineering, University of Isfahan, Isfahan, Iran

*Rabbani.h@ieee.org

**Abstract**

Diabetic Retinopathy (DR) caused by diabetes occurs as a result of changes in the retinal vessels and causes visual impairment. Microaneurysms (MAs) are the early clinical signs of DR, whose timely diagnosis can help detecting DR in the early stages of its development. It has been observed that MAs are more common in the inner retinal layers compared to the outer retinal layers in eyes suffering from DR. Optical Coherence Tomography (OCT) is a non-invasive imaging technique that provides a cross-sectional view of the retina and it has been used in recent years to diagnose many eye diseases. As a result, in this paper has attempted to identify areas with MA from normal areas of the retina using OCT images. This work is done using the dataset collected from FA and OCT images of 20 patients with DR. In this regard, firstly Fluorescein Angiography (FA) and OCT images were registered. Then the MA and normal areas were separated and the features of each of these areas were extracted using the Bag of Features (BOF) approach with Speeded-Up Robust Feature (SURF) descriptor. Finally, the classification process was performed using a multilayer perceptron network. For each of the criteria of accuracy, sensitivity, specificity, and precision, the obtained results were 96.33%, 97.33%, 95.4%, and 95.28%, respectively. Utilizing OCT images to detect MAs automatically is a new idea and the results obtained as preliminary research in this field are promising.

## I. INTRODUCTION

Diabetic Retinopathy (DR) is a serious and dangerous disease and since the prevalence of DR is directly related to the increasing prevalence of diabetes, it is growing rapidly in different societies [1]. DR can damage the retinal blood vessels and ultimately leads to blindness. So, early detection is very important. Microaneurysms (MAs) are the first clinical signs of DR [2] and their premature diagnosis can be very useful for preserving patients' vision. Retinal MAs are small protrusions of the capillary vessel walls and are most commonly seen in the inner nuclear layer [3-5]. The number of MAs indicates the likelihood of developing a more severe level of DR [6]. The Fluorescein Angiography (FA) imaging technique has been used for over 50 years as the gold standard technique for retinovascular imaging [7]. Ophthalmologists currently use FA as the primary tool for the detection of MAs in the images of DR patients. Although FA is capable of detecting microvasculature details, it is an invasive and time-consuming method that requires intravenous injection and an expert photographer [8]. The imaging technique known as Optical Coherence Tomography (OCT) was introduced in 1991 [9] and subsequently became an essential tool for clinical imaging [8]. OCT is a near-infrared light interferometry method developed for cross-sectional, noninvasive, high-resolution and three-dimensional tomography imaging in biological systems. Therefore, the use of OCT images as a new type of retinal imaging for automatic detection of MAs could be a good approach to improve the current routine of MA detection. In this project, previous research related to the detection of microaneurysms is reviewed. Previous articles have commonly used FA or fundus imaging techniques to investigate the characteristics of MAs and their detection, and the following results have been obtained: MAs are often presented as hyperfluorescent points on FA images, and some MAs are associated with focal fluorescence leakage. Initially, MA counting protocols were developed to monitor the progress of DR during drug trials. Automated techniques were then used for more accurate and faster MA detection [10-14]. Although FA shows a very clear image of the retina and many MAs are detectable in these images, it is not well accepted by patients because of its invasive nature. Therefore, the researchers then used fundus images, which are a less invasive method than FA to detect MAs. Most of the methods available for the detection of MAs using fundus images are performed in two stages including: 1) extraction of MA candidates and 2) classification. The first step requires image pre-processing to reduce the noise and improve the contrast. Subsequently, candidate regions for MAs are identified. Blood vessel segmentation algorithms are then used to eliminate the blood vessels from the MA candidates to reduce false positives, since many blood vessels may appear as false positives in the pre-processed images. Secondly, the classification algorithm is applied to the classification of these features to the candidates of MA and the candidate of non-MA [15]. In the recent works [16-26], the same procedure has been applied by employing different machine learning, deep learning and image processing methods. However, processing of fundus images is more challenging than FAs because different objects in fundus images may be confused with diabetes lesions. In addition, the major limitation of these images is that they provide a two-dimensional view of the 3D retinal texture [27]. So nowadays, attention is being drawn to OCT imaging, as a new imaging technique.

Automatic detection of ocular diseases using OCT images is still in its infancy because only academic research has been published and no commercial activity is available [28]. Since OCT images facilitate the evaluation of retinal morphology to microscopic resolution [12], it has been attempted to use these images to detect various retinal malformations such as glaucoma, Diabetic Macular Edema (DME), and Age-related Macular Degeneration (AMD), Retinal detachment, Diabetic retinopathy [29]. Most previous works have focused on the analysis of OCT images on the problem of dividing the retinal layers, Retinal thickness measurement, or segmentation of specific lesions such as cysts [28]. OCT is more desirable than fundus imaging because it provides in-depth and cross-sectional information from the eye [30].

In recent years, researchers have investigated the potential of using SD-OCT images to detect MAs and have attempted to identify the apparent features of diabetic MAs in these images. In these studies [31-35], the following results have been achieved for appearance of MAs:

MAs have a relatively circular and capsular structure and mainly located in the Inner Nuclear Layer (INL) layer and the deeper layers than the INL layer. MAs increase the thickness of the retina, and in some cases reflectivity points are found in their vicinity.

Preliminary studies have shown that SD-OCT is a reliable and promising criterion for further investigation to diagnose MAs and also has a good correlation with the equivalent FA image [36]. Therefore, in this paper, we intended to develop a system for the first time to automatically detect MAs using OCT images. For this purpose, first, the desired database including FA and OCT images of patients with DR disease was collected. Then registration of OCT B-scans was performed with corresponding FAs and MAs and normal areas from B-scan images were extracted. In the next step, the characteristics of each of the areas related to the two categories, i.e., MA and normal, were extracted using the BOF method, and finally these characteristics were used to train the multilayer perceptron network, so that the two regions could be separated.

The remainder of the paper is organized in the following order. The methods and algorithms used in this paper are described in Section II, the experimental results are reported in Section III and the discussions and conclusions are expressed in Section IV.

## II. METHOD

The method proposed in this paper for automatic detection of MAs in OCT images is divided into two general stages of system training and testing (Fig. 1). Each of these steps consists of several sections, which will be described in more detail below.

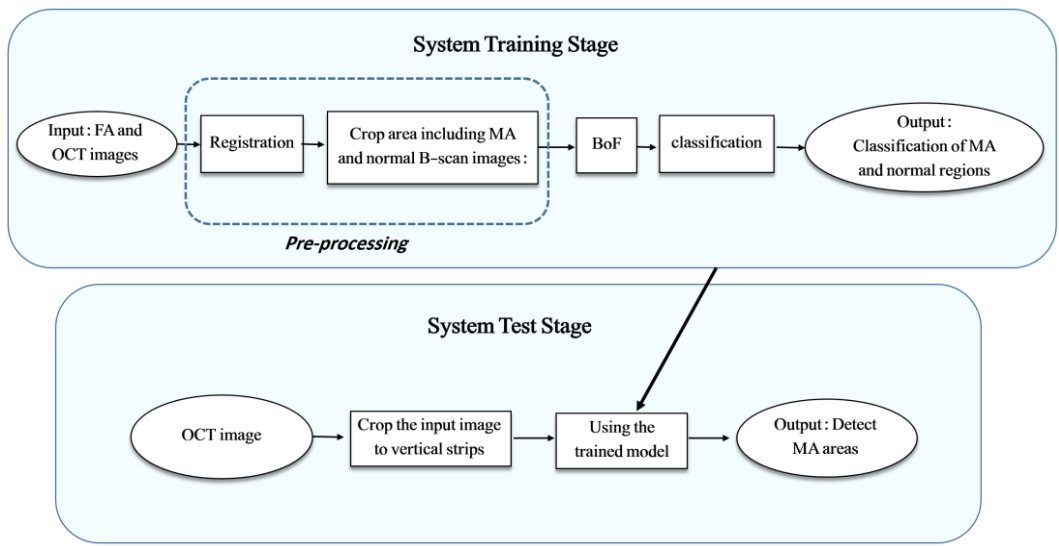

Fig. 1. The general diagram of the method used in this study.

FA and OCT images used in this study were captured using the Spectral HRA2 / OCT Heidelberg imaging device. For this study, late frames of FA imaging of 20 patients with DR were used. Equivalent to each FA image, 31 OCT B-scan images were produced by the imaging unit and each B-scan dimension was 496 *768 pixels.

*A. System training*

In the training phase of the system, the FA images and their equivalent OCT images are initially considered as inputs of the system. The following are the three steps of system training:

1) *Pre-processing*

As it is shown in Fig. 1, the preprocessor consists of two parts: 1) registration of OCT B-scans with corresponding FAs and 2) extraction of MA and non-MA areas from B-scan images.

*a) Registration:* Given that patients have DR, MAs are visible on FA images of each patient. After registration of FA and OCT images, the location of MAs on OCT images will be specified. Since OCT images provide a cross-sectional and depth view of the retina and the FA images represent a surface view of the retina, the two images have different dimensions and cannot be directly aligned together [37]. As stated, the dataset used in this study was prepared by the Heidelberg device, which enables simultaneous Scanning Laser Ophthalmoscopy (SLO) / OCT imaging. SLO provides real-time images from the surface of the retina which is corresponding to the OCT images. Since these two image groups (SLO and OCT) are made from a unit light source, they are pixel-to-pixel aligned. So, by registration of SLO and FA images, OCT B-scans and FAs will be aligned. For this reason, the multi-step correlation-based technique has been used for the registration of FA and OCT images [38]. This algorithm consists of two phases: 1) Phase I: general registration (rigid analysis) and 2) Phase II: local registration in areas with MA (non-rigid analysis) which are briefly explained below.

Phase I: First, the veins network is extracted from each image using the Dijkstra Forest Exploratory Algorithm used in [39], and the correlation of FA and SLO images is calculated using a multi-step correlation-based method at different angles and dimensions. Then the most appropriate affine parameters are identified with the highest correlation as a result of the algorithm and used in subsequent phases.

Phase II: A patch-based local registration algorithm with several different accuracies is used to improve the results of the first phase of registration in areas with MA. After this step, MA areas can be identified in B-Scan OCT images. The result of the registration and the MA regions identified on the FA image and its equivalent B-Scan for an example set of the images used in this study are shown in Fig. 2.

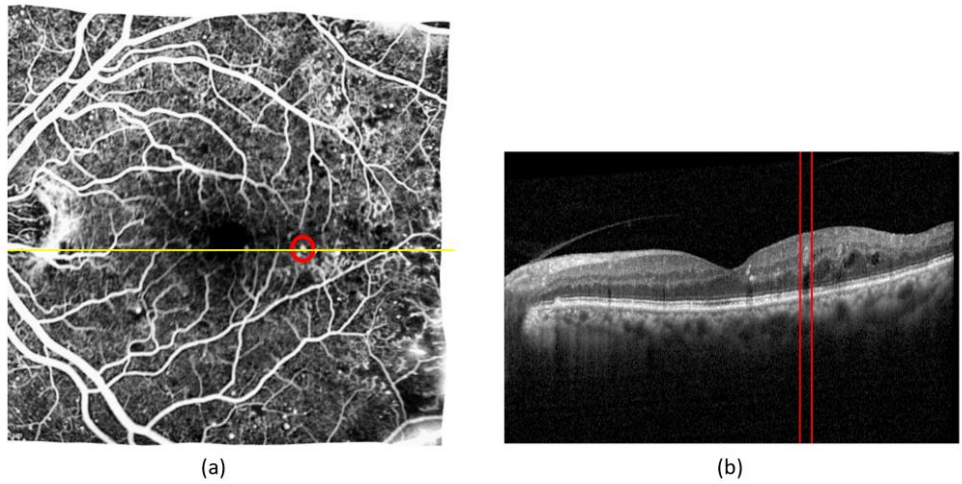

Fig. 2. The FA image (a) and the B-scan image corresponding to the yellow horizontal line displayed on the FA image (b). The location of MA is shown as a red circle in (a) image and red lines in (b) image.

Extraction of MA and non-MA areas from B-scan images: After identifying the MAs in FA images and applying the registration stage, the corresponding areas can be extracted from the B-scan images (the red strip in Fig. 2.b). A range of 30 pixels is considered for each detected MA in the B-scan image and unnecessary areas are removed from the top and bottom of the region of interest (ROI) to preserve useful information around the retinal layers. Therefore, for each desired area, the vertical bar consists of 170*30 pixels. Normal areas will similarly be extracted from the B-scan image. Fig. 3 shows two areas extracted from the MA and normal areas from a sample B-scan image. In total, 92 MA areas and 110 normal areas were used for this study. Some areas were excluded from the MA category due to other lesions related to other diseases. 70% of data are randomly selected for training, 15% for test and 15% for validation.

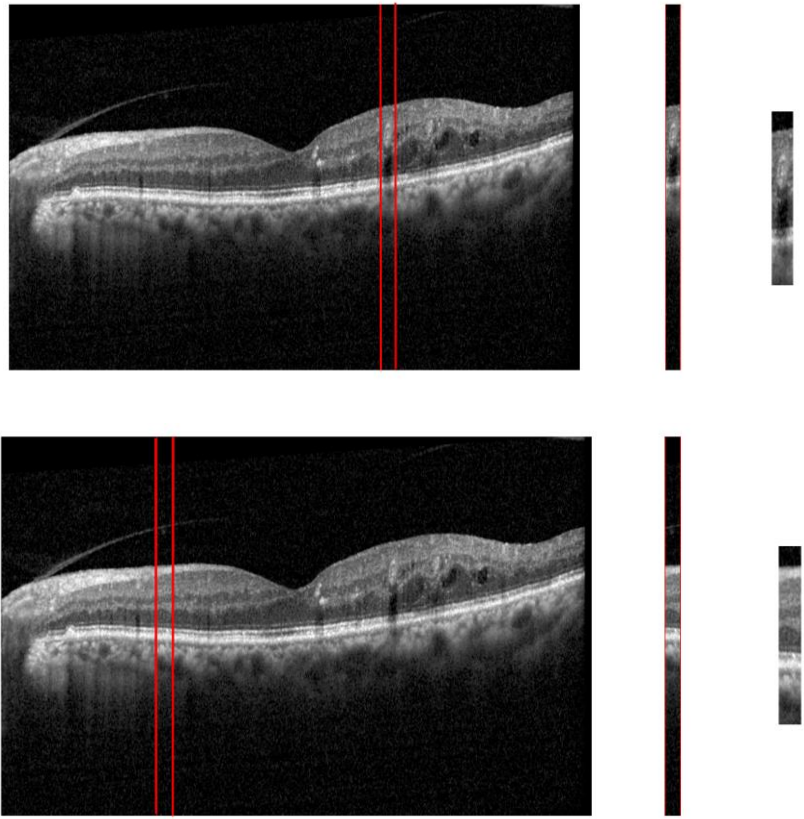

Fig. 3.Extraction of MA (top) and normal (bottom) areas from a B-scan image. In both cases, the middle image is an area of 20 pixels extracted from the original image, and in the right image, unnecessary parts are removed from the top and bottom.

2) *The Bag of Features*

*a) Feature extraction:* In this study, SURF was used to identify and describe image features. The dense SURF local descriptors have been used to identify and describe areas with more useful information. For this purpose, key SURF points were extracted from image patches to describe the distribution of local light intensity. This enables local descriptors to identify and localize key points in the horizontal and vertical directions of image patches. Since these features are invariant to scale and rotation and are resistant to noise [40], small MAs with low contrast are also detected. Fig. 4 shows points with an [8 x 8] grid in a sample image of the dataset used in this study. As it can be seen, according to the SURF algorithm for describing key points, the square area aligned with the direction of the point is regularly subdivided into small 4*4 pixels square subareas, which makes important spatial information around the point to be preserved. Then, Haar wavelet responses are obtained for each subarea, where according to Fig. 4, *dx* is the Haar wavelet response in the horizontal direction and *dy* is the Haar wavelet response in the vertical direction. So for each part, the sum of *dx*, | *dx* |, *dy,* and | *dy* | is considered as the descriptor vector.

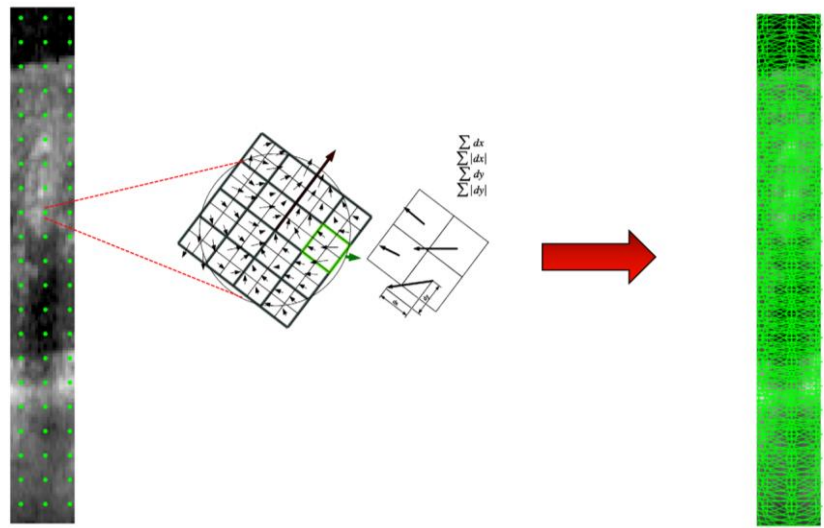

Fig. 4. Point descriptors produced by SURF. On the left image, the key points of a sample image with an [8 x 8] grid are shown, and how the SURF descriptor vector is produced, is shown in the middle of the figure.

*b) Quantization and building visual vocabulary:* In the visual words construction phase, the number of features extracted from the previous stage will be reduced since these features become vectors of key points for bags through histograms and in this method the feature vectors created by SURF must be grouped into several clusters by a clustering algorithm. In this study, the K-means++ clustering algorithm with the Euclidean distance criterion is used. The reason for choosing this clustering method is its simplicity. The K-means++ clustering algorithm performs better than the original K-means method by changing the way of the initialization of the clusters' centers [41]. In other words, an encryption method is used to count the number of visual words that occur in an image based on the center of the clusters.

The sum of the events of each visual word (each cluster) for all images in each category is shown in Fig. 5. These two histograms show the difference between the two categories of images in their occurrence rate for 100 visual words.

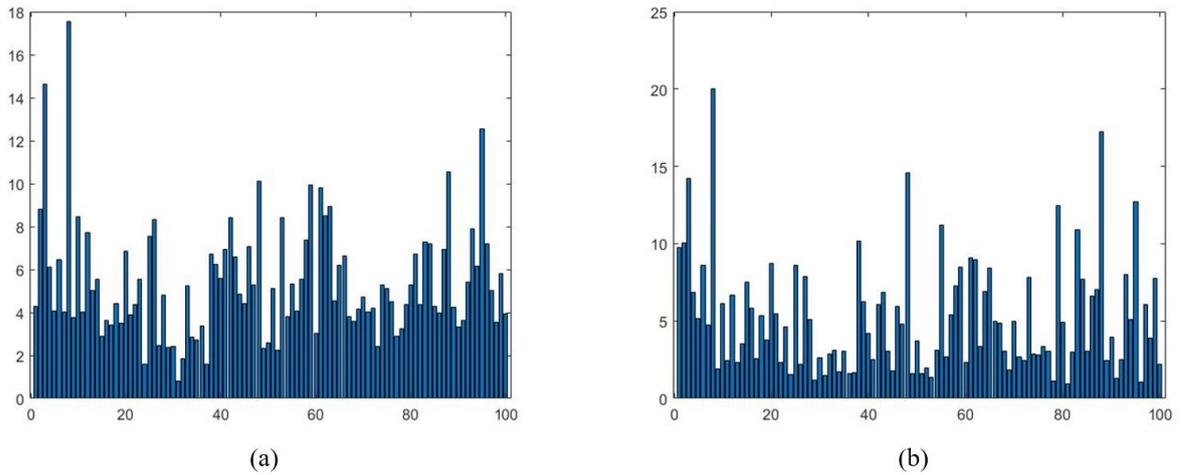

Fig. 5. Sum of events of all visual words in each category. The image (a) is for the MA class and the image (b) is for the normal class.

3) *Classification by a multilayer perceptron network*

Visual word-based image classification is a process in which a training set of images are given to the classifier for training in the form of term vectors. In this study, a multilayer perceptron (MLP) network with a hidden layer, and 10 neurons were used for the classification process. By changing the number of hidden layer neurons, the multilayer perceptron network accuracy criterion was examined and according to the result of this evaluation, which can be seen in Fig. 6, using 10 neurons are the most appropriate choice. In the experiments performed, the MLP classification method showed better performance than other classification algorithms such as Gaussian SVM, Linear SVM, KNN, and Naïve Bayes. The results are shown in TABLE I.

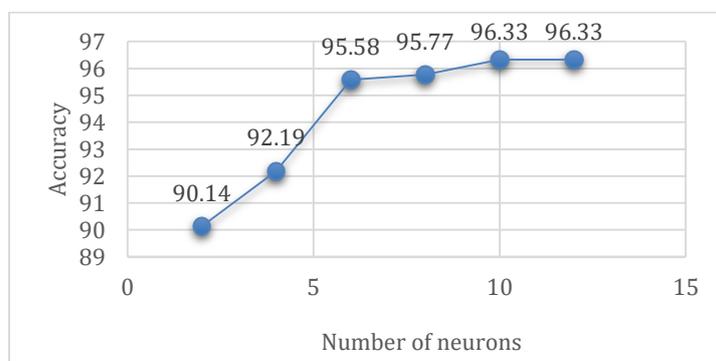

Fig. 6. The sensitivity analysis. The effect of the number of hidden layer neurons of the multilayer perceptron network on the accuracy criterion

*B. System testing*

In the testing phase, the following steps will be taken for each OCT image as input to identify the MAs:

1) Extraction of the vertical strips from each B-scan image

2) Encoding each extracted strip using the dictionary created in the training step. That way, first the features of each strip are extracted and then assigned to the nearest centers of the clusters. The number of each term that appears in the image represents a normal histogram (the term vector).

3) This term vector is given to the MLP network trained in the previous step to assign a label to it.

4) The predicted label is compared to the actual label. This comparison will show the accuracy of the system in detecting MAs.

### III. EXPERIMENTAL RESULTS

To evaluate the proposed method, several classification methods were tested in comparison to a multilayer perceptron network. The comparison of these methods for the criteria of accuracy, sensitivity, specificity, and precision are presented in TABLE I.

Besides, the impact of the existence of a BOF technique to achieve the desired result was evaluated. So instead of using BOF, the image features were first extracted using SURF. Then PCA is applied to reduce a large number of features. Finally, these feature vectors are given as input to different classifiers including MLP, Gaussian SVM, Linear SVM, KNN, and Naïve Bayes. The average of the measured criteria for these classifiers is shown in Table I. These results show a decrease in accuracy compared to when using BOF.

Also in Fig.7, a bar chart is plotted for each criterion to better compare the methods described in Table I.

TABLE I. EVALUATION OF THE PROPOSED METHOD

| | classification method | evaluation criteria | | | |
|---|---|---|---|---|---|
| | | Accuracy | Sensitivity | Specificity | Precision |
| 1 | BOF+MLP | 96.33% | 97.33% | 95.4% | 95.28% |
| 2 | BOF+Gaussian SVM | 91.63% | 88.27% | 95.02% | 94.65% |
| 3 | BOF+Linear SVM | 91.13% | 86.52% | 95.78% | 95.40% |
| 4 | BOF+KNN | 84.19% | 73.14% | 95.28% | 93.95% |
| 5 | BOF+Naïve Bayes | 87.75% | 85.65% | 89.91% | 89.48% |
| 6 | PCA+MLP | 92.52% | 93.93% | 92.22% | 92.36% |
| 7 | PCA+ Gaussian SVM | 83.38% | 85.53% | 81.28% | 82.16% |
| 8 | PCA+ Linear SVM | 80.94% | 77.77% | 84.15% | 83.19% |
| 9 | PCA+KNN | 57.32% | 14.77% | 99.39% | 96.32% |
| 10 | PCA+ Naïve Bayes | 76.13% | 76.14% | 76.15% | 76.14% |

Based on these results, the use of MLP as the classification algorithm in this method yields the best results. Then, the SVM classifier with a Gaussian kernel comes in the second.

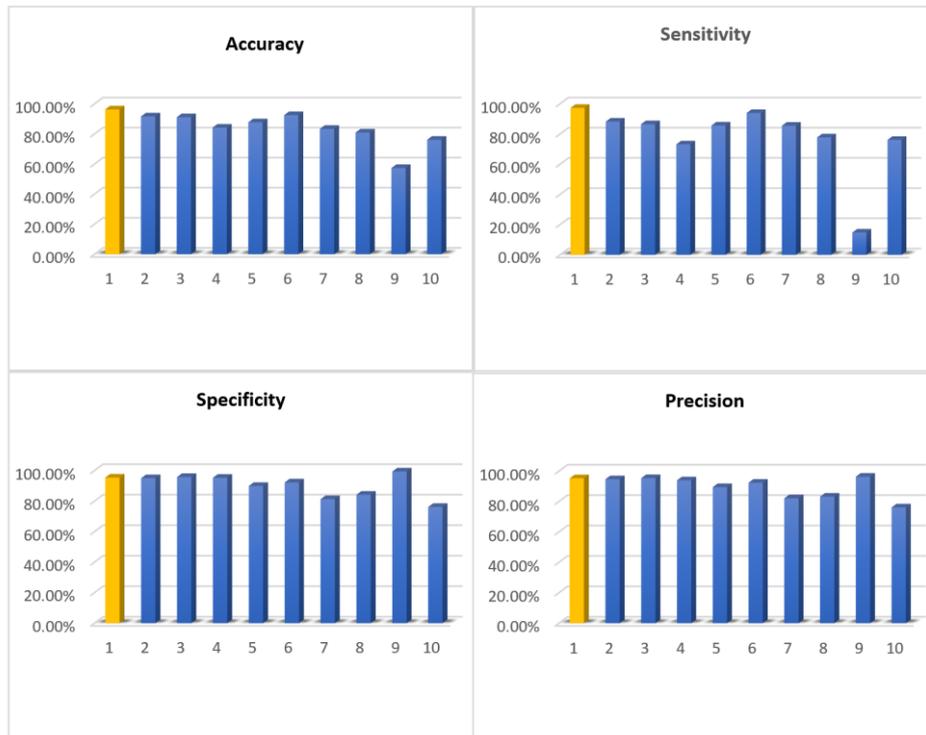

Fig. 7. The charts related to 4 criteria of accuracy, sensitivity, specificity and precision. The numbers in the x-axis of each chart is corresponding to the proposed methods in TABLE I.

# IV. DISCUSSIONS AND CONCLUSIONS

Given that diabetes is a rapidly growing disease, the traditional diagnosis of DR is very time-consuming, labor cost, and requires expert people. Therefore, a system for screening and monitoring diabetic patients and automatic diagnosis of DR would be beneficial. Since MAs are the first clinical signs of DR, their diagnosis can help early diagnosis of DR. In the published ratings for DR, the difference between the two levels without DR and Mild Non-proliferative DR is the presence or absence of MA. Therefore, if there are only MAs in an image and the system fails to detect them, the patient is diagnosed mistakenly without DR, causing the disease to become more severe and elevated to a high level in the next visit. This demonstrates the importance of automatic MA detection algorithms and their accuracy and sensitivity.

In the Introduction section of this article, the superior applicability of OCT as a screening method and its advantages over fundus photography in DR screening courtesy of its strength in providing in-depth information from the retina, have been explored. Therefore, in this article, it was decided to OCT imaging over fundus photography.

It should be noted that OCTA imaging is also a new imaging technique that has become popular in recent years. The 3D visualization provided by OCTA images is useful for analyzing MAs and their spatial orientation [3]. MAs appear as saccular or fusiform in these images [42]. A number of previous studies have compared FA and OCTA imaging and concluded that FA detects a higher number of MAs than OCTA [43-46]. Measurement of relative blood speed indicates that blood flow is slower in MAs [47], [48]. Studies have also identified MAs without blood flow. As a result, fewer MAs are detected in OCTA images because OCTA is based on blood flow [45] and areas where blood flow is slow may not be seen in these images [48]. Studies show that OCTA is more effective in diagnosis of MAs than fundus Imaging [49]. On the other hand, as described in the text, in OCT images, MAs are seen as a reflection. If these reflections are hyporeflective, they will less likely visible in OCTA images. But if they are hyperreflective, they will often be visible in OCTA images [50].

Due to the above reasons, it was decided to use OCT images for automatic detection of MAs. For this purpose, FA images were first registered with OCT B-scans. By doing so, the MAs regions are identified in OCT B-scans and can be analyzed. SURF has been used to extract the features that can detect precise local information from the MA regions and distinguish areas containing MA and normal well. Visual words were then created using the BOF method. Utilizing the BOF method significantly improves the results. Finally, a multilayer perceptron network was applied to classify these areas using these visual words. Eventually, the evaluation criteria show suitable performance for the method proposed in this study. In this paper, it was shown that in addition to identifying areas with clear signs of MA, even areas that do not have clear signs of MA and are not visible to the human eye in the related OCT B-scans can be identified with the help of this method. These areas can be identified with the help of features extracted from OCT images and machine learning algorithms.

This research is a preliminary study to show that OCT images can be used to detect MAs. However, this method is not without its caveats. Limitations of this method include: Lack of suitable and comprehensive data for this purpose and incorrect diagnosis of MAs in some cases due to the presence of lesions caused by other diseases and blood vessels in OCT images. To overcome these limitations, the following solutions can be considered in future works.1) Expansion and improvement of datasets, 2) Using prior and posterior B-scan information for each B-scan to increase accuracy and prevent the possible loss of an MA, and 3) Expanding the categories from two categories: "MA" and "normal" to four categories. "MA", "non-MA",

"normal", and "vessel". So that the "MA" category only includes areas related to MA without any other lesions and vessels. The "non-MA" category includes areas that do not have MA but may have other lesions. The "normal" category includes areas without any lesions or vessels. Finally, the class of "vessel" refers to the areas that only contain the vessel.

In addition to the above and as stated in [51], MAs can also be examined in terms of flow, location and capillary density. Thus, although OCTA devices are not as common as OCT devices in all medical centers, they may be potentially helpful and worthy of further research for MA analysis.